# Solving a maze with a quantum computer

Mark A. Bashuk*
*Center for Advanced Studies in Math, Science, and Technology, Wheeler High School,
Marietta, GA 30067*
(January 17, 2003)

It is well known, and appreciated, that quantum computers have the potential to be the most powerful computational devices ever created. This newfound power comes from a 'quantum parallelism' effect that allows the computer to be in multiple states at the same time. This property of quantum parallelism, while suited to handle common problems such as factoring and searching an unorganized database, is extremely well-suited to handle the task of solving a binary maze. I propose an algorithm that can be used to solve a binary maze on a quantum computer, with guaranteed accuracy. While it does work, it does come with a few setbacks, in that the maze must have no flaws, and that the computer requires a number of qubits equal to the number of decisions in the maze, plus $\log_2$ of the decisions.

Quantum computers have the potential to be the most powerful computational devices ever created [1], and while there are nearly infinite ways to use them effectively, only a few are of academic or industrial use [2]. It is known that quantum computers are fast at searching databases, and that they are even faster at factoring prime numbers [3-4]. What I am proposing is a simple, yet effective algorithm that by using minimal resources can solve any binary maze much faster than any classical computer.

A binary maze is a simple thing. It is the common tree-branching shape accompanying any single-elimination bracket [5]. But the most important thing about a binary maze, especially for my algorithm, is that there are no exceptions to the design. Each line has one input, and two outputs, until we reach the end of the "maze" in which there is no further output. This rule has zero exceptions, as any violations of it would cause utter chaos in a world where chaos is enough of a problem already. This action of one input, and two outputs, causes a huge problem for classical computers. Because it doubles in size each time, the problem takes exponentially more time to solve as the number of decisions increases [5]. For example, a 100 step binary maze would have 2^100 decisions at the end of the maze, a number too large to be easily dealt with.

Solving mazes, while not a tough task for an 8-year old, is not difficult because of the simplicity of the maze. Given a maze of large enough size, a computer would be needed to solve the maze quickly. While there are many algorithms for solving a maze, the most noted and useful algorithm for solving a maze of any type is called depth-first search [6]. Depth-first search works by exhausting every branch of the maze, while attempting to limit the number of tries. Depth-first search is equivalent to walking down a hallway, keeping your right hand on the wall, and only turning around at each dead-end, keeping track of where you go. This method gets the job done quicker than any other non-probabilistic algorithm could, but given a large enough problem, it is intractable [6].

Given this problem, a beautiful and simple answer arises from the chaos. Use a quantum computer. A quantum computer can try all the paths of a binary maze at one time, and dramatically reduce the time needed to solve a childish problem like a maze [7].

What I propose is a simple algorithm for solving a maze with a quantum computer. A few of the characteristics must be pre-determined for this algorithm to work. First, the binary maze must have no defects in it; that is all of the inputs except the last must have two outputs. The last line will not have any outputs, because otherwise it would not be the last line. Also, there will be a *number of qubits* in the quantum computer equal to the number of the decisions in the maze, plus $\log_2 N$, with N being the number of decisions in the maze.

The first step in the maze is the most obvious. The only reason the state of the first gate matters is because of the counting qubits. With a maze of N-decisions, we will have $\log_2 N + N$ qubits in the computer. The first $\log_2 N$ qubits (the counting qubits) are for keeping track of our position in the maze, which is what each gate reads to know what part of the qubit to change. Besides that, the rest of the zeros are just used as reference so that there is an obvious starting qubit.

The reason $\log_2 N$ qubits are used for the number of counting qubits is that because with N decisions, you need $\log_2 N$ binary digits to keep room for every possible number [8]. Therefore, step 2 of the algorithm, reads the first $\log_2 N$ qubits (the counting qubits), and prepares to move to that position.

After reading the counting qubits and determining our position, we move to that qubit and perform our transformation on it. At each decision, we have a choice of going up or going down, and that is represented with a 1 or a 0. The advantage of the quantum computer is that we can simultaneously travel up and down at the same time [1]. Along those lines, the 1s and 0s of the qubit represent each possible decision. Step 3 puts the qubit into a superposition of up and down at the point that step 2 declared.

Step 4 is the simplest of all the steps, but still quite necessary. Step 4 changes the counting qubits by one, so that the next gate knows to move along in the maze. That is the small clause that the extra $\log_2 N$ qubits allows us to do. Otherwise, it would be a guessing game as to what qubits to modify. Step 5 tells us to repeat the process described in steps 2, 3, and 4 until we reach the "end" of the maze. The end of the maze is the last row in the maze, in which there are no decision to be made. We have just traversed through N decisions, and being composed of binary decisions, there are $2^N$ possibilities at the end of the maze.

Step 6 completes the algorithm. Of the many points at the end of the maze, the creator determines one point to be the end. It is at this point that we look at the value of the qubit that is in the computer. The order of the 1s and 0s in the qubit tells us the path that leads to the corresponding endpoint. At first glance, this seems silly, as knowing the end point of the maze is what appears we are trying to find. But, if we try to solve a normal maze, we are always given a beginning and an ending point, and our job is to find the path in between. My algorithm does exactly that.

While this algorithm does work, there are several problems associated with it. The applications of many quantum algorithms are instantaneously obvious [2]. Unfortunately for mine, no non maze-related applications seem feasible. While there could be some applications developed in the future, that is a much more difficult problem than creating the algorithm. Also, the accuracy required in discerning the solution to a complicated maze is extremely high. On a large maze, possibly N=100 decisions, there would be 2^100 different possibilities, and the differences between many of them is just a few parts in 100. While that is the problem for the engineer of the quantum computer running this algorithm, it does provide a definite deterrent to implementation.

In comparison to depth-first search on a classical computer, this algorithm is exponentially faster. In a worst-case scenario, depth-first search would take 59 steps to solve a maze requiring five decisions [9]. A quantum computer could do it in five steps. Even on average, the difference of thirty to three is quite large. The difference becomes much more influential as the size of the maze increases; a maze of 50 decisions would require $1.125 \times 10^{16}$ steps, while a quantum computer could do it in fifty, a percentage difference of $4.4 \times 10^{-12}$%. A modern computer can perform $10^9$ calculations per second, meaning it would take it 13 days to solve what a quantum computer could solve in seconds.

As Vandersypen, et. al. noted, experimental realization of quantum algorithms can already be done on five qubit NMR quantum computers [10]. While a quantum computer with that few qubits could be used to solve a two-step maze, a computer of large enough size to solve a maze of difficulty has yet to be created. Hopefully, by the time I am able to pursue my own endeavors in this field, either massively qubit quantum computers will be built, or new algorithms will be designed to take advantage of my process.
In conclusion, I have presented a completely new quantum algorithm that can be used to solve a maze. While it is currently only a theory, a quantum computer of large enough size needs to be built to test this out. My algorithm works with guaranteed accuracy, if the experimenter has enough accuracy on his measurement tools, and the maze is designed in accordance with my definition.

I thank D.R. Finkelstein for mentoring me and introducing me to this field. I must thank Mohsen Shiri, M. Chapman and T.A.B. Kennedy of GT for their hospitality. I would like to thank Jeff Rosen, Chris Neill, David Berta, Rad Fraasa, Ryan Kane, and everyone else at Wheeler with whom I discussed this. I would like to thank Gregg Farmer, Cathie Banks, Phyllis Boudreaux, and Pam Diers for the opportunities of my internship.

* Email Address: CokeisBomb@aol.com.